\documentclass[journal=jacsat,manuscript=article]{achemso}

\usepackage[version=3]{mhchem} 
\usepackage{braket}

\author{Daniel F. Calero-Osorio}
\author{Paul W. Ayers}
\email{ayers@mcmaster.ca} 

\affiliation[McMaster University]
{Department of Chemistry, McMaster University, Hamilton, Ontario L8S 4M1, Canada}

\title[High-truncation SZ-CT]
  {Seniority-Zero Canonical Transformation Theory: Error Reduction Via Late Truncation}

\abbreviations{}

\SectionNumbersOn
\begin{document}

\begin{abstract}
We show how to add the effects of residual electron correlation to a reference seniority-zero wavefunction by transforming the true electronic Hamiltonian into seniority-zero form. The transformation is treated via the Baker–Campbell–Hausdorff (BCH) expansion and the seniority-zero structure of the reference is exploited to evaluate the first three commutators exactly; the remaining contributions are handled with a recursive commutator approximation, as is typical in canonical transformation methods. By choosing a seniority-zero reference and using parallel computation, this method is practical for small- to medium-sized systems. Numerical tests show high accuracy, with errors $\sim 10^{-4}$ Hartree.
\end{abstract}

\section{Introduction}
The accurate description of strongly correlated electronic systems remains a challenging problem in quantum chemistry, partly because achieving quantitative accuracy requires treating the multireference nature of strong/static electron correlation (which typically entails high, even exponential, computational cost), but also because the interplay between dynamic and static correlation must be modeled. Generally, the static correlation is modeled
using complete active space self-consistent field (CASSCF)\cite{ROOS1980157, siegbahn1981complete,  roos1987complete, schmidt1998construction, szalay2012multiconfiguration}, complete active space configuration interaction (CASCI) \cite{slavivcek2010ab,szalay2012multiconfiguration,levine2021cas,fales2017complete} or multiconfigurational self-consistent field (MCSCF)\cite{RUEDENBERG198241,doi:https://doi.org/10.1002/9780470142943.ch7, 10.1143/PTPS.40.37,Wahl1977} computations. Dynamic correlation is usually added with configuration interaction (CI), \cite{doi:https://doi.org/10.1002/9781119019572.ch11,szabo1996modernCI,Lowdin1955,KnowlesHandy1984,OlsenRoos1988,SherrillSchaefer1999,ShavittBartlett2009,SzalayMuller2012} coupled-cluster (CC) ,\cite{ShavittBartlett2009,doi:https://doi.org/10.1002/9781119019572.ch13,szabo1996modernCC,Coester1960,Cizek1966,paldus1982relationship,CrawfordSchaefer2000,BartlettMusial2007} and many-body perturbation theories (MBPT).\cite{ShavittBartlett2009,PhysRev.46.618,doi:https://doi.org/10.1002/9781119019572.ch14,szabo1996modernMBPT,MollerPlesset1934,Goldstone1957,Bloch1958,FetterWalecka1971,LindgrenMorrison1985} 

For processes (e.g., bond-breaking) and systems (e.g., polyradicals) with strong electron correlation, both static correlation and dynamic correlation must be modeled. These combined methods generally start with a multireference wave function that describes static correlation (typically CAS) and then add dynamic correlation, though
dynamic-then-static\cite{liuIdeasRelativisticQuantum2010a,durandDirectDeterminationEffective1983,lindgrenCoupledclusterApproachManybody1978}
and static-then-dynamic-then-static\cite{huangIVIIterativeVector2017,leiFurtherDevelopmentSDSPT22017,liuICIIterativeCI2016a,
liuSDSStaticdynamicstaticFramework2014a,zhangFurtherDevelopmentICIPT22021} strategies have also been explored. Examples of the conventional static-then-dynamic strategy are multireference Moller-Plesset (MRPT) \cite{HIRAO1992374,Hirao1993,WolinskiPulay,Wolinski,Nakano1993,GrimmeWaletzke2000,witek2002intruder} , N-electron valence perturbation theory (NEVPT)\cite{ANGELI2001297,10.1063/5.0051211,10.1063/1.1361246,Angeli2001b,angeli2002n,Angeli2004a}, and complete active space perturbation theory (CASPT) \cite{Roos1982,doi:10.1021/j100377a012,Andersson1992,Andersson1995,Forsberg1997,Finley1998,Ghigo2004}. Most of the multireference perturbation theories (MRPT) are known to suffer from intruder states, where near degenerancies causes divergencies in the perturbative expansions; they are also significantly more computationally demanding than their single-reference counterparts. Intruder-states free MRPT like NEVPT2 offer a solution for that problem, with the downside of storing and evaluating up to four-body reduced density matrices (RDMs), which is only available for small active spaces. Non-perturbative approaches include multireferece configuration interaction (MRCI) \cite{10.1063/1.455556,10.1063/1.439365,buenker1974individualized} and multireference coupled cluster \cite{JeziorskiMonkhorst1981,RittbyBartlett1991,PaldusPiecuchPylypowJeziorski1993,mahapatra1999size,lyakh2012multireference,kohn2013state}, which offer higher accuracy than the MRPT methods but incur problems like poor scalability, redundancies, etc.\cite{ivanov2011state,https://doi.org/10.1002/jcc.540141112,ivanov2011state, stampfuss1999parallel}.

Although CAS wave functions generally suffice to describe static correlation, selecting the correct active space requires intuition, though automated approaches can work for straightforward cases or when including excess orbitals in the active space is not considered cost-prohibitive.\cite{kaufoldAutomatedActiveSpace2023,
kellerSelectionActiveSpaces2015,
kingRankedOrbitalApproachSelect2021,
sayfutyarovaAutomatedConstructionMolecular2017,
steinAutomatedSelectionActive2016,
tothComparisonMethodsActive2020} Doubly-occupied configuration interaction (DOCI; also called closed-shell or seniority-zero CI) offers an alternative, where instead of restricting the space of orbitals one restricts oneself to electron configurations where all electrons are paired. Dating back at least to the 1960s, DOCI and related seniority-zero methods typically give excellent results for strong correlation, at least when the strong correlation is associated with the breaking of chemical bonds.\cite{bytautasSeniorityNumberDescription2015,
alcobaHybridConfigurationInteraction2014a,
veillardCompleteMulticonfigurationSelfconsistent1967,
cookDoublyoccupiedOrbitalMCSCF1975,
carboGeneralMulticonfigurationalPaired1977,CaleroOsorio2025SZ} 
The seniority-zero restriction reduces the size of the Hilbert space by a power of 1/2, but the cost of evaluating DOCI remains factorial. This motivates the search for efficient parameterizations of the DOCI wavefunction,\cite{CaleroOsorio2025SZ} a quest that even predates the first DOCI work.\cite{hurley1953molecular, parks1958theory} The most popular strategy is to represent the DOCI wavefunction as an antisymmetric product of geminals.\cite{Surjn1999,DEBAERDEMACKER2024185}. Examples include the anti-symmetric product of strongly orthogonal geminals (APSG),\cite{surjanIntroductionTheoryGeminals1999,rassolov2002geminal,surjan2012strongly,parr1956generalized,parks1958theory,kutzelnigg1964direct,jeszenszki2015local,pernal2013equivalence} generalized valence bond perfect-pairing (GVB-PP),\cite{hurley1953molecular, parks1958theory, hunt1972self, hay1972generalized, small2014coupled,lawler2008symmetry,cullen1996generalized,moss1975generalized,dykstra1980perfect,carter1988correlation,hartke1992ab} the antisymmetrized product of 1-reference orbital geminals (AP1roG),\cite{doi:10.1021/ct300902c,PhysRevB.89.201106,10.1063/1.4880819,Boguslawski2015} and Richardson/rank-2 geminals.\cite{JOHNSON2013101,10.1063/5.0088602,faribaultReducedDensityMatrices2022,
fecteauNearexactTreatmentSeniorityzero2022,
fecteauReducedDensityMatrices2020,
johnsonBivariationalPrincipleAntisymmetrized2022,
johnsonRichardsonGaudinMeanfield2020,
johnsonRichardsonGaudinStates2024,
tecmerAssessingAccuracyNew2014a}
For systems with more than 2 electrons, DOCI and its parameterizations can only describe static electron correlation. This motivates strategies for treating dynamic correlation by either enhancing the parameterization (e.g., non-seniority-zero geminal theories\cite{JOHNSON2013101,johnsonStrategiesExtendingGeminalbased2017b,johnsonRichardsonGaudinStates2025,
johnsonRichardsonGaudinStatesNonzero2025,johnsonSingletGeminalWavefunctions2024}) or adding corrections for dynamic correlation using configuration interaction, perturbation theory, density functional theory, or coupled-cluster approaches.\cite{lawler2008symmetry, cullen1996generalized,moss1975generalized,carter1988correlation,huUnitaryBlockCorrelatedCoupled2025,liBlockcorrelatedCoupledCluster2004,robb1970generalized,robb1971generalized,robb1972generalized,limacher2014simple, 10.1063/1.4906607,garza2015range,Boguslawski2015,tecmer2022geminal,CaleroOsorio2025SZ} 

Our strategy is philosophically different: we ask whether we can transform the electronic Hamiltonian into seniority-zero form, so that DOCI wavefunctions would be \textit{exact} eigenfunctions of the transformed Hamiltonian.\cite{CaleroOsorio2025SZ} This approach is clearly linked to strategies that use unitary/similarity transformations to add dynamic correlation to multireference wave functions.\cite{10.1063/1.5094643,10.1063/1.4890660,doi:10.1021/acs.jctc.5b00134,10.1063/1.4947218,10.1063/5.0059362} However, our inspiration was canonical transformation (CT) theory for Hamiltonians.\cite{10.1063/1.3086932,10.1063/1.2196410,C2CP23767A,Neuscamman01042010} In CT, the Baker-Campbell-Hausdorff (BCH) formula is used to evaluate a unitary transformation of the Hamiltonian but, instead of truncating the expansion, terms are estimated by using operator decomposition to approximate many-body operators in terms of one- and two-body operators and terms in the BCH expansion are included until they become negligible. CT theory is an affordable approach for including dynamic correlation into multireference wave functions, but relies on strong assumptions. First, the reference wavefunction (modeling static correlation) must be sufficiently close to the exact wave function for the operator decomposition process to be justified. This condition cannot be always met, and in those cases, to improve the accuracy of the method high-order excitation operators and their associated reduced density matrices (RDMs) need to be included. However, three- and four-body RDMs are accessible only for small- to medium-sized active spaces (when CT is used to add dynamic correlation to CASSCF/CASCI). 

Instead of seeking a tranformation that downfolds the full electronic Hamiltonian to an active space, we seek a transformation to a seniority-zero Hamiltonian.\cite{CaleroOsorio2025SZ} Because of the pairing structure inherent in seniority-zero wave functions, the evaluation of 3- and 4-RDMs becomes computationally tractable, allowing late truncation of the BCH expansion. 
The remainder of this paper is organized as follows. Sec. \ref{Theory} reviews CT theory and seniority-zero reduced density matrices, including the spin-free (Section \ref{spinfree}) formalism we use. Section \ref{implementation} describes the the general aspects of our computational implementation, including the strategies we use to include many-body operators and RDMs without undue computational cost (Sec. \ref{cost_reduction}). In Sec. \ref{results} we apply the method to three molecules: \ce{H8} in the STO-6G basis, \ce{BH} in the 6-31G and cc-pVDZ basis sets, and \ce{N2} in the STO-6G basis. Finally, Sec. \ref{conclusions} offers our conclusions and outlines perspectives for future work.

\section{Theory}
\label{Theory}
\subsection{Late truncation in seniority-zero canonical transformation theory (LT-SZCT)}
\label{LT-SZCT}
As usual in CT theory, we start with a (seniority-zero) reference wave function, $\ket{\Psi_0}$, and add the missing dynamic correlation using a unitary transformation:
\begin{equation}
\begin{aligned}
\ket{\Psi} = e^{\hat{A}}\ket{\Psi_0},
\end{aligned}
    \label{eq1}
\end{equation}
where $\hat{A}$ is an anti-hermitian operator constructed from excitation and de-excitation operators:
\begin{equation}
\begin{aligned}
\hat{A} =\sum_{p,q}a_{pq}\left(\hat{E}^{p}_{q} - \hat{E}^{q}_{p} \right) +\frac{1}{2}\sum_{p,q,r,s}a_{pqrs}\left(\hat{E}^{pq}_{rs} -\hat{E}^{rs}_{pq} \right),
\end{aligned}
    \label{eq2}
\end{equation}
where $a_{pq}$ and $a_{pqrs}$ are the generator's one- and two-body amplitudes, and $a_{pqrs}$ is anti-symmetric with respect to individual interchanges of upper/lower indices. We choose to work in the spin-free formalism in which $\hat{E}^{p_1p_2p_3,...,p_n}_{q_1q_2q_3,...,q_n}=\hat{E}^{\dagger}_{p_1}\hat{E}^{\dagger}_{p_2}\hat{E}^{\dagger}_{p_3}...\hat{E}^{\dagger}_{p_n}\hat{E}_{q_n}\hat{E}_{q_{n-1}}...\hat{E}_{q_1}$, represent products of creation $\hat{E}^{\dagger}_{p}$ and annihilation $\hat{E}_{q}$ spin-free operators in the spatial orbitals $p$ and $q$ respectively. The generic indices $p,q,r,s$ run over all spatial orbitals. The appropriate generator $\hat{A}$ that maps the (seniority-zero) reference to the system's wave function is found by variational optimization:
\begin{equation}
\begin{aligned}
E=\min_{A}&\bra{\Psi_0}e^{-\hat{A}}\hat{H}e^{\hat{A}}\ket{\Psi_0}.
\end{aligned}
    \label{eq3}
\end{equation}
This expression would be exact if we did not truncate $\hat{A}$ after 2-body terms and, instead, evaluated the BCH expansion exactly.\cite{10.1063/1.5133059,doi:10.1021/acs.jpca.5b02015} In contrast to CT theory, we decided to use variational optimization instead of the generalized Brillouin conditions (GBCs) to determine the generator $\hat{A}$. In general, for a small enough generator we expect both approaches to give similar results, however, the evaluation of the GBCs in CT theory imposes extra constraints on the generator because the Jacobian is approximated by discarding all terms $\mathcal{O}(A)$ and larger. Thus, our variational optimization will differ from the CT's GBCs-based solutions in cases where $\mathcal{O}(A)$ and $\mathcal{O}(A^2)$ terms in the BCH expansion are nonnegligible.

To evaluate the unitary transformation we use the Baker–Campbell–Hausdorff (BCH) expression:
\begin{equation}
\begin{aligned}
e^{-\hat{A}}\hat{H}e^{\hat{A}} = \hat{H} &+ \left[\hat{H},\hat{A}\right] + \frac{1}{2!} \left[\left[\hat{H},\hat{A}\right],\hat{A}\right]  \\[8pt] &+ \frac{1}{3!} \left[\left[\left[\hat{H},\hat{A}\right],\hat{A}\right], \hat{A}\right] +...,
\end{aligned}
    \label{eq4}
\end{equation}
If we could evaluate the unitary transformation exactly, Eq. (\ref{eq3}) would be an upper bound to the true ground-state energy. However, given the structure of the generator, the BCH expansion does not truncate. Moreover, each additional commutator adds an higher-order interaction term. E.g., if $\hat{H}$ and $\hat{A}$ are two-body operators, then $[\hat{H},\hat{A}]$ includes three-body operators and $\left[[\hat{H},\hat{A}],\hat{A} \right]$ includes four-body operators. Therefore, to evaluate the BCH expansion to $\mathcal{O}\left(\lVert\hat{A}\rVert^2\right)$ one needs access to the 3- and 4-electron reduced density matrices (RDMs), which are expensive to compute and store in traditional methods. However, by choosing a seniority-zero wave function as our reference, we are able to compute and store the 4RDM with the same computational cost/memory scaling as a regular, non-seniority-zero, 2RDM (see section \ref{sz-reference}). The rest of the terms need to be approximated. We approximate the fourth- and higher-order terms using the operator decomposition expressions introduced in canonical transformation theory (CT), wherein each commutator $[\hat{H},\hat{A}]$ will be approximated by one- and two-electron operators and RDMS: $[\hat{H},\hat{A}]_{1,2}$. Thus, the expression we will use to evaluate the unitary transformation is: 
\begin{equation}
\begin{aligned}
e^{-\hat{A}}\hat{H}e^{\hat{A}} \approx \hat{H} &+ \left[\hat{H},\hat{A}\right] + \frac{1}{2!} \left[\left[\hat{H},\hat{A}\right],\hat{A}\right]  \\[8pt] &+ \frac{1}{3!} \left[\left[\left[\hat{H},\hat{A}\right]_{1,2},\hat{A}\right]_{1,2}, \hat{A}\right]_{1,2} +...~.
\end{aligned}
    \label{eq5}
\end{equation}
As mentioned before, starting with the third term we are making approximations, so the transformation is not exactly unitary and the variational minimization (eq \ref{eq3}) is not an upper bound for the energy. However, when the size of the generator  $\hat{A}$ is small enough, higher-order terms in Eq. \ref{eq5} are small, the errors incurred by the operator decomposition are tolerable, and the BCH expansion converges rapidly. 

\subsection{\label{spinfree}Operator decomposition for spin-free operators}
We refer the reader to the literature for a thorough explanation of the spin-free operator decomposition;\cite{C2CP23767A,Kutzelnigg10022010,10.1063/1.3256237} here we present only the key results needed by our approach.  
The spin-free creation and annhilation operators are defined by tracing over the spin degrees of freedom of the standard spin-orbital creation/annihilation operators,
\begin{equation}
\begin{aligned}
    E^{p_1}_{q_1}&=\sum_{\sigma= \alpha, \beta}a^{\dagger}_{p_1 \sigma}a_{q_1 \sigma}, \\[8pt]
    E^{p_1 p_2}_{q_1 q_2}=&\sum_{\sigma,\tau= \alpha, \beta}a^{\dagger}_{p_1 \sigma}a^{\dagger}_{p_2 \tau}a_{q_2 \tau}a_{q_1 \sigma},\\[8pt]
    E^{p_1 p_2 p_3}_{q_1 q_2 q_3}=&\sum_{\sigma,\tau, \nu= \alpha, \beta}a^{\dagger}_{p_1 \sigma}a^{\dagger}_{p_2 \tau}a^{\dagger}_{p_3 \nu}a_{q_3 \nu}a_{q_2 \tau}a_{q_1 \sigma},
\end{aligned}
    \label{eq6}
\end{equation}
and the corresponding spin-free reduced density matrices are:
\begin{equation}
\begin{aligned}
    \Gamma^{p_1}_{q_1} &=\langle\Psi |E^{p_1}_{q_1} |\Psi\rangle,\\[8pt]
    \Gamma^{p_1 p_2}_{q_1 q_2}&=\langle\Psi|  E^{p_1 p_2}_{q_1 q_2}|\Psi\rangle,\\[8pt]
    \Gamma^{p_1 p_2 p_3}_{q_1 q_2 q_3}&=\langle\Psi| E^{p_1 p_2 p_3}_{q_1 q_2 q_3} |\Psi\rangle.
\end{aligned}
    \label{eq7}
\end{equation}
The spin-free operator decomposition equations are defined using the generalized normal order (GNO) formalism\cite{Kutzelnigg10022010}, wherein we normal order a particular excitation operator $\hat{o}$ with respect to a (possibly multi-configurational) wave function $\ket{\Psi}$. The expressions for the GNO one-, two-, and three-body spin-free excitation operators are:
\begin{equation}
\begin{aligned}
    \tilde{E}^{p_1}_{q_1}&={E}^{p_1}_{q_1} - {\Gamma}^{p_1}_{q_1},\\[8pt]
    \tilde{E}^{p_1 p_2}_{q_1 q_2}&= {E}^{p_1 p_2}_{q_1 q_2} - \sum\left(-\frac{1}{2}\right)^{x} {\Gamma}^{p_1}_{q_1}\tilde{E}^{p_2}_{q_2} -{\Gamma}^{p_1 p_2}_{q_1 q_2},\\[8pt]
    \tilde{E}^{p_1 p_2 p_3}_{q_1 q_2 q_3}  &= {E}^{p_1 p_2 p_3}_{q_1 q_2 q_3} - \sum\left(-\frac{1}{2}\right)^{x}{\Gamma}^{p_1}_{q_1}\tilde{E}^{p_2 p_3}_{q_2 q_3} \\[5pt] &- \sum\left(-\frac{1}{2}\right)^{x}{\Gamma}^{p_1 p_2}_{q_1 q_2} \tilde{E}^{p_3}_{q_3} - \Gamma^{p_1 p_2 p_3}_{q_1 q_2 q_3},
\end{aligned}
    \label{eq8}
\end{equation}
where the notation $\sum(-1)^xA^{p_1 p_2...p_k}_{q_1 q_2 .... q_k}B^{p_{k+1} p_{k+2}..}_{q_{k+1} q_{k+2} .... q_k}$  means that there is one term for each partition of the indices ${p_i},{q_i}$ among the objects $A$ and $B$, where the ${p_i}$ are kept on top and the ${q_i}$ are kept on bottom, and a minus $\frac{1}{2}$ factor is applied for each permutation that violates the original pairing.\par
When we re-write the commutator $[\hat{H},\hat{A}]$ as $[\hat{H},\hat{A}]_{1,2}$, we are implicitly assuming that our reference wavefunction is accurate enough so that the unitarily-transformed Hamiltonian, $\bar{H}=e^{\hat{A}}\hat{H}e^{-\hat{A}}$, can be accurately approximated by a two body operator. I.e., we assume that the expectation value of the three-body spin-free GNO operator, $ \tilde{E}^{p_1 p_2 p_3}_{q_1 q_2 q_3}$, is negligible. 
Then, we express the three-body excitation operator in terms of one- and two-body operators and RDMs:
\begin{equation}
\begin{aligned}
    {E}^{p_1 p_2 p_3}_{q_1 q_2 q_3}  &=  \sum\left(-\frac{1}{2}\right)^{x}{\Gamma}^{p_1}_{q_1}E^{p_2 p_3}_{q_2 q_3} + \sum\left(-\frac{1}{2}\right)^{x}{\Gamma}^{p_1 p_2}_{q_1 q_2}E^{p_3}_{q_3} \\[8pt] 
    &-2 \sum\left(-\frac{1}{2}\right)^{x}{\Gamma}^{p_1}_{q_1}{\Gamma}^{p_2}_{q_2}E^{p_3}_{q_3} + \sum\left(-\frac{1}{2}\right)^{x}{\Gamma}^{p_1}_{q_1}{\Gamma}^{p_2}_{q_3}E^{p_3}_{q_2}  \\[8pt] 
    &+2 \sum\left(-\frac{1}{2}\right)^{x}{\Gamma}^{p_1}_{q_1}{\Gamma}^{p_2}_{q_2}\Gamma^{p_3}_{q_3} -\sum\left(-\frac{1}{2}\right)^{x}{\Gamma}^{p_1}_{q_1}{\Gamma}^{p_2}_{q_3}\Gamma^{p_3}_{q_2} \\[8pt]
    &-2 \sum\left(-\frac{1}{2}\right)^{x}{\Gamma}^{p_1}_{q_1}\Gamma^{p_2 p_3}_{q_2 q_3} + \Gamma^{p_1 p_2 p_3}_{q_1 q_2 q_3}.
\end{aligned}
    \label{eq9}
\end{equation}
Typically the 3RDM is decomposed in terms of the one- and two-RDMS using a cumulant decomposition. However, for a seniority-zero reference wave function, computing, storing, and operating with the 3RDM is completely tractable. The last equation is used iteratively in the BCH expansion to re-write everything past the third term in terms of one and two-body excitation operators and up to three-body RDMs. 
Note that the last equation is completely independent of the choice of reference wave function. As we shall see in the next section, in our case, due to the structure of the seniority-zero reference wave function, the one-, two-, and three-body RDM will not have two-, four-, and six-indices but one-, two-, and three-indices, respectively.

\subsection{Seniority-zero Reduced density matrices}
\label{sz-reference}
For seniority-zero wave functions, the only non-zero elements in reduced density matrices (of any order) correspond to excitations that preserve the number of paired electrons in the wave function. This means that the number of indices needed to describe the $k$-RDM of a seniority-zero wavefunction is half the number that is required for a general wavefunction. The 1RDM is thus diagonal:
\begin{equation}
\Gamma^{p}_p= \langle\Psi_{SZ}    |\hat{c}^{\dagger}_{p}\hat{c}_{p}|\Psi_{SZ}\rangle.
    \label{Eq10}
\end{equation}
Non-diagonal elements are zero since they change the occupancies per orbital.\par
The two-electron RDM requires two indices and has two non-zero blocks:
\begin{equation}
\begin{aligned}
    \Gamma^{p\bar{p}}_{q\bar{q}} &= \langle\Psi_{SZ}    |\hat{c}^{\dagger}_{p}\hat{c}^{\dagger}_{\bar{p}}\hat{c}_{q}\hat{c}_{\bar{q}}|\Psi_{SZ}\rangle, \\
    \Gamma^{pq}_{pq} &= \langle\Psi_{SZ}|\hat{c}^{\dagger}_{p}\hat{c}^{\dagger}_{q}\hat{c}_{p}\hat{c}_{q}|\Psi_{SZ}\rangle,
\end{aligned}
    \label{Eq11}
\end{equation}
where  $p$, $\bar{p}$ refer to electrons in the same spatial orbital with different spin.
The elements $\Gamma^{p\bar{p}}_{q\bar{q}}$ are called pair-correlation terms while the elements $\Gamma^{pq}_{pq}$ are called diagonal elements. The two blocks above capture the usual non-zero index patterns of the seniority-zero 2RDM, but they are not exhaustive: symmetry-related counterparts must also be included. For the 2RDM, this is simple, the only additional block is $\Gamma^{pq}_{qp}$. Recall that since we are working in the spin-free frame, $\Gamma^{pq}_{qp} \neq -\Gamma^{pq}_{pq}$. \par
The 3RDM of a general wave function is a six-index tensor but for a seniority-zero wavefunction one has three nonzero blocks with at most three indices:
\begin{equation}
\begin{aligned}
\Gamma^{pqr}_{pqr} &=  \langle\Psi_{SZ}|\hat{c}^{\dagger}_{p}\hat{c}^{\dagger}_{q}c^{\dagger}_{r}\hat{c}_{p}\hat{c}_{q}\hat{c}_{r}|\Psi_{SZ}\rangle,     \\[5pt]
 \Gamma^{pq\bar{q}}_{pq\bar{q}} &=  \langle\Psi_{SZ}|\hat{c}^{\dagger}_{p}\hat{c}^{\dagger}_{q}c^{\dagger}_{\bar{q}}\hat{c}_{p}\hat{c}_{q}\hat{c}_{\bar{q}}|\Psi_{SZ}\rangle,     \\[5pt]  
 \Gamma^{pq\bar{q}}_{pr\bar{r}} &= \langle\Psi_{SZ}|\hat{c}^{\dagger}_{p}\hat{c}^{\dagger}_{q}\hat{c}^{\dagger}_{\bar{q}}\hat{c}_{p}\hat{c}_{r}\hat{c}_{\bar{r}}|\Psi_{SZ}\rangle. 
 \label{eq12}
\end{aligned}
\end{equation}
Notice that there are elements in the 3RDM with only two indices, which are already contained in the 2RDM. Including symmetry operations, we end up with 24 non-zero blocks. However, only 15 blocks are three-index blocks. As we'll show in section \ref{cost_reduction}, the number of blocks can be significantly reduced using the symmetry relations between blocks.

A similar result is obtained for the 4RDM.
Normally, this object contains eight indices, but for seniority-zero wave functions there are six blocks with at most 4 indices, so its complexity is similar to the 2RDM of general (non-seniority-zero) wave functions. The nonzero 4RDM bocks are:
\begin{equation}
\begin{aligned}
\Gamma^{pqrs}_{pqrs} &=  \langle\Psi_{SZ}|\hat{c}^{\dagger}_{p}\hat{c}^{\dagger}_{q}\hat{c}^{\dagger}_{r}c^{\dagger}_{s}\hat{c}_{p}\hat{c}_{q}\hat{c}_{r}\hat{c}_{s}|\Psi_{SZ}\rangle,    \\[5pt]
 \Gamma^{p\bar{p}qr}_{p\bar{p}qr} &=  \langle\Psi_{SZ}|\hat{c}^{\dagger}_{p}\hat{c}^{\dagger}_{\bar{p}}\hat{c}^{\dagger}_{q}c^{\dagger}_{r}\hat{c}_{p}\hat{c}_{\bar{p}}\hat{c}_{q}\hat{c}_{r}|\Psi_{SZ}\rangle,       \\[5pt]  
 \Gamma^{p\bar{p}q\bar{q}}_{p\bar{p}q\bar{q}} &= \langle\Psi_{SZ}|\hat{c}^{\dagger}_{p}\hat{c}^{\dagger}_{\bar{p}}\hat{c}^{\dagger}_{q}\hat{c}^{\dagger}_{\bar{q}}\hat{c}_{p}\hat{c}_{\bar{p}}\hat{c}_{q}\hat{c}_{\bar{q}}|\Psi_{SZ}\rangle,      \\[5pt]  
 \Gamma^{p\bar{p}r\bar{r}}_{p\bar{p}s\bar{s}} &= \langle\Psi_{SZ}|\hat{c}^{\dagger}_{p}\hat{c}^{\dagger}_{\bar{p}}\hat{c}^{\dagger}_{r}\hat{c}^{\dagger}_{\bar{r}}\hat{c}_{p}\hat{c}_{\bar{p}}\hat{c}_{s}\hat{c}_{\bar{s}}|\Psi_{SZ}\rangle, 
  \\[5pt]  
 \Gamma^{pqr\bar{r}}_{pqs\bar{s}} &= \langle\Psi_{SZ}|\hat{c}^{\dagger}_{p}\hat{c}^{\dagger}_{q}\hat{c}^{\dagger}_{r}\hat{c}^{\dagger}_{\bar{r}}\hat{c}_{p}\hat{c}_{q}\hat{c}_{s}\hat{c}_{\bar{s}}|\Psi_{SZ}\rangle, 
   \\[5pt]  
 \Gamma^{p\bar{p}r\bar{r}}_{q\bar{q}s\bar{s}} &= \langle\Psi_{SZ}|\hat{c}^{\dagger}_{p}\hat{c}^{\dagger}_{\bar{p}}\hat{c}^{\dagger}_{r}\hat{c}^{\dagger}_{\bar{r}}\hat{c}_{q}\hat{c}_{\bar{q}}\hat{c}_{s}\hat{c}_{\bar{s}}|\Psi_{SZ}\rangle.  \\
 \label{eq13}
\end{aligned}
\end{equation}
After adding symmetry operations, there are 177 nonzero blocks of the 4-RDM. 

\section{Implementation}
\label{implementation}
Starting with the one- and two-electron integrals,\cite{10.1063/5.0006074,sunLibcintEfficientGeneral2015,
kimGBasisPythonLibrary2024}, we used a development version of \texttt{PyCI} and \texttt{HORTON 3} to perform an orbital-optimized DOCI calculation.\cite{chanTaleHORTONLessons2024,10.1063/5.0219010}
We obtained the symbolic formula for the commutators $[\hat{H},\hat{A}]$, $\left[[\hat{H},\hat{A}],\hat{A}  \right]$ and $[\hat{H},\hat{A}]_{1,2}$ using the \texttt{sqa} software package \cite{10.1063/1.3086932}
The numerical evaluation of the tensor contractions in the symbolic formula is performed using \texttt{Numpy} \texttt{einsum} and \texttt{opt\_einsum},\cite{daniel2018opt} which suffices for prototyping.
The seniority-zero RDMs of the DOCI wave function were obtained by extending \texttt{PyCI} to include higher-order and seniority-zero RDMs. Finally, the energy is computed by tracing the transformed electron integrals against the RDMs:
\begin{equation}
    \begin{aligned}
     E= \sum_{pq}\bar{h}_{pq}\Gamma_{pq} +\frac{1}{2} & \sum_{pqrs}\bar{v}_{pqrs}\Gamma_{pqrs} +\frac{1}{6}\sum_{pqrstu}\bar{v}_{pqrstu}\Gamma_{pqrstu} \\
     &+\frac{1}{24}\sum_{pqrstuvw}\bar{v}_{pqrstuvw}\Gamma_{pqrstuvw},
     \label{eq14}
    \end{aligned}
\end{equation}
where the new set of electron integrals ($\bar{h}_{pq}$, $\bar{v}_{pqrs}$, $\bar{v}_{pqrstu}$, $\bar{v}_{pqrstuvw}$) are functions of the generator $\hat{A}$. Finally, we pass the function that computes the energy to a \texttt{Scipy} minimizer, together with a constraint over the norm of the generator $\hat{A}$, that is specific for each molecule configuration. The initial guess for the generator parameters is the zero vector (identity transformation); small random perturbations around zero lead to the same converged energies within numerical tolerance. Near convergence, we observed that \texttt{trust-constr} may perform additional ``polishing'' iterations in which the energy remains unchanged to numerical precision while optimality/feasibility measures are refined. Therefore, we impose a conservative maximum-iteration cap after the energy plateaus; this does not affect the reported energies.

In the present implementation, the one- and two-body generators entering $\hat{A}$ are used in their natural spin-free form and are not explicitly orthogonalized. Because our approach employs a single seniority-zero reference state and a purely variational determination of the generator amplitudes, we do not encounter the severe redundancy and multiple parentage problems characteristic of CAS/MRCC formulations with an explicit core-active-virtual partition. Operator-level quasi-redundancy manifests as flat directions in the variational landscape (near-zero curvature modes) rather than as numerical instabilities or multiple solutions of a projected amplitude system. In our tests this only affected the number of quasi-Newton iterations required for convergence and did not change the final energies within numerical precision. An explicit orthogonalization of the generator manifold, along the lines of existing canonical transformation and MRCC schemes, could further improve the conditioning of the optimization and is a natural avenue for future refinements of this method.

The selection of the constraint on the magnitude of the generator is based on intuition (e.g., how far the DOCI reference is from the actual FCI curve, or in other words, how much dynamic correlation needs to be added).
We are exploring strategies for automatically determining the maximum value for the norm of the generator $\hat{A}$, so that the transformation is quasi-unitary and the LT-SZCT energy is always an upper bound to the exact energy.
We note that this type of tuning parameter is quite common in multireference methods and that our results are insensitive to (reasonable) restrictions on the generator's norm.

\subsection{Computational details}
\label{cost_reduction}
The computational cost of the method is determined by three main operations. First, the evaluation of the commutators $[\hat{H},\hat{A}]$, $\left[[\hat{H},\hat{A}],\hat{A}\right]$, and the truncated commutator $[\hat{H},\hat{A}]_{1,2}$. The two commutators scale as $\mathcal{O}(N^5)$; the truncated commutator $[\hat{H},\hat{A}]_{1,2}$, obtained via the operator decomposition, scales as $\mathcal{O}(N^7)$. Second, the energy evaluation, which involves tracing the transformed electron integrals with the reference RDMs, scales as $\mathcal{O}(N^8)$. Finally, due to the antisymmetry of the two-body amplitudes of the generator $A$, the total number of independent elements of the gradient scales as $\mathcal{O}\left(\frac{N^2 (N-1)^2}{4}\right)$, which for less than approximately 150 spatial orbitals is effectively  $\mathcal{O}(N^3)$. Thus, for our target applications we say the gradient evaluation scales effectively as $\mathcal{O}(N^3)$. Therefore, without any assumptions about the reference wave function or further code optimizations, the overall scaling of the method would be $\mathcal{O}(N^{11})$.

By employing seniority-zero (SZ) wave functions as the reference, significant optimizations can be achieved. As discussed in Section \ref{sz-reference}, the SZ RDMs contain far fewer nonzero elements than typical RDMs, which we exploit in two ways:
\begin{enumerate}
    \item Instead of contracting over all possible index combinations, we explicitly extract the nonzero elements of the SZ RDMs and contract only over them. For example, rather than computing the energy using Eq.~\ref{eq14}, we evaluate it as:
\begin{equation}
    \begin{aligned}
     E= \sum_{p}\bar{h}_{pp}\Gamma^{SZ}_{pp} &+\frac{1}{2}  \sum_{P \in SZ}\sum_{pq}\bar{v}_{P\{pqpq\}}\Gamma^{SZ}_{P\{pqpq\}}\\ &+ \frac{1}{6}\sum_{P \in SZ}\sum_{pqr}\bar{v}_{P\{pqrpqr\}}\Gamma^{SZ}_{P\{pqrpqr\}} \\
     &+\frac{1}{24}\sum_{P \in SZ }\sum_{pqrs}\bar{v}_{P\{pqrspqrs\}}\Gamma_{P\{pqrspqrs\}},
     \label{eq15}
    \end{aligned}
\end{equation}
where, in the notation $\sum_{P \in SZ}\sum_{pq}\bar{v}_{P\{pqpq\}}\Gamma^{SZ}_{P\{pqpq\}}$, the first summation over $P \in SZ$ denotes the set of index permutations $(p,q)$ that generate a nonzero block of the SZ RDMs. For the 2RDM, there are three such nonzero blocks: $\Gamma^{SZ}_{pqpq}$, $\Gamma^{SZ}_{ppqq}$, and $\Gamma^{SZ}_{pqqp}$. For the 3RDM, the number increases to 15, corresponding to the three blocks introduced in the previous section together with all symmetry-related permutations of the indices. For the 4RDM, this results in 177 nonzero blocks. Although this strategy increases the total number of tensor contractions, it drastically reduces their computational scaling. Comparing Eq.~\ref{eq14} with Eq.~\ref{eq15}, the cost of evaluating the contractions is reduced from $\mathcal{O}(N^8)$ to $\mathcal{O}(N^4)$.
\item We explicitly map the nonzero blocks of the full RDMs to new tensors with reduced dimension. Thus, instead of working with the full four-index tensor representing the 2RDM, we work directly with its three nonzero two-index blocks:
\begin{equation}
    \begin{aligned}
\Gamma^{SZ}_{ppqq} \Rightarrow \Gamma^{2_{ppqq}}_{pq}, \\[10pt]
\Gamma^{SZ}_{pqpq} \Rightarrow \Gamma^{2_{pqpq}}_{pq}, \\[10pt]
\Gamma^{SZ}_{pqqp} \Rightarrow \Gamma^{2_{pqqp}}_{pq},
     \label{eq16}
    \end{aligned}
\end{equation}
where the tensors $\Gamma^{2_{ppqq}}_{pq}$ only have two indices. We apply the same procedure for the 3- and 4-RDMs. Thus, instead of storing and operating on the full six- and eight-index tensors (3RDM, 4RDM), we only manipulate tensors with at most four indices. This strategy does not reduce the formal scaling, since the number of free indices in the contractions remains the same, but it significantly improves the actual computation time because manipulating high-rank tensors is inherently inefficient: even simple operations, such as multiplying an entire tensor by a constant, require iterating over the full structure. This leads to poor cache performance and substantial overhead in memory transfers, since large contiguous blocks of memory need to be moved or accessed repeatedly. By remapping the 8-index tensor into several lower-index tensors, the memory footprint of each tensor is smaller and the data can be accessed more efficiently. 
\end{enumerate}
\

Additional code optimization was made by computing all the contributions from the three- and four-body transformed electron integrals $\bar{v}_{pqrstu}$ and $\bar{v}_{pqrstuvw}$ on the fly, so that these tensors are never actually generated and stored. Finally, the analytical gradient of the transformation is implemented and parallelized, such that at the end the actual scaling of the gradient computation becomes $\mathcal{O}(\frac{N^3}{n_c})$, where $n_c$ is the number of cores available for the calculation. With these optimizations, the scaling of the truncated commutator $[\hat{H},\hat{A}]_{1,2}$ becomes $\mathcal{O}(N^5)$ and the evaluation of the energy $\mathcal{O}(N^4)$; the ``effective'' scaling of the method is $\mathcal{O}(\frac{N^8}{n_c})$. 

\subsection{Formal comparison between SZ-LCT and LT-SZCT}
In previous work,\cite{caleroosorio2025seniorityzerolinearcanonicaltransformation}  we presented the seniority-zero linear canonical transformation theory (SZ-LCT), which like LT-SZCT transforms a molecular Hamiltonian $\hat{H}$ into a seniority-zero Hamiltonian $\hat{H}_{SZ}$ using unitary transformations. The two methods use different objective functions and different approximations to the BCH expansion. While the present method minimizes the energy using the variational minimization of Equation \ref{eq3}, SZ-LCT minimizes the non-seniority-zero elements of the transformed Hamiltonian:
\begin{equation}
\begin{aligned}
&\min_{\hat{A}}\left(|| \hat{\widetilde{H}}_{\text{non}-SZ}||\right), \\[10pt]
&\hat{\widetilde{H}}_{\text{non}-SZ}= \hat{\widetilde{H}}-\hat{\widetilde{H}}_{SZ},
\label{eq17}
\end{aligned}
\end{equation}
where $\hat{\widetilde{H}}_{SZ}$ is the seniority-zero sector of the transformed Hamiltonian $\hat{\widetilde{H}}$. Thus, while LT-SZCT is effectively transforming the electronic Hamiltonian into a seniority-zero Hamiltonian by eliminating the contribution to the energy of the non-seniority-zero elements, SZ-LCT is directly minimizing those matrix elements. 

The second distinction between the two methods lies in the evaluation of the unitary exponential and, specifically, in the truncation strategy adopted for the BCH expansion. SZ-LCT employs the standard commutator approximation of canonical transformation theory, in which all contributions to the BCH series are represented by operators truncated at the two-body level. This approximation introduces a truncation error of $\mathcal{O}(||A||)$ in the evaluation of the transformation. In contrast, LT-SZCT exploits the structure of the seniority-zero RDMs to evaluate the single and double commutators ($[\hat{H},\hat{A}],[[\hat{H},\hat{A}],\hat{A}]$) exactly, so the truncation error is of $\mathcal{O}(||A||^3)$. Remarkably, LT-SZCT has the same computational scaling as SZ-LCT. Late truncation is particularly advantageous when the constraints on the generator $\hat{A}$ are relaxed during the minimization, as it improves the stability of the energy curve and enables the treatment of regimes where a larger generator is required to recover missing correlation---conditions under which SZ-LCT may yield less reliable results.

\section{Results}
\label{results}
\subsection{\ce{H_8}} 
\begin{figure}[h!]
\includegraphics[width=0.7\linewidth]{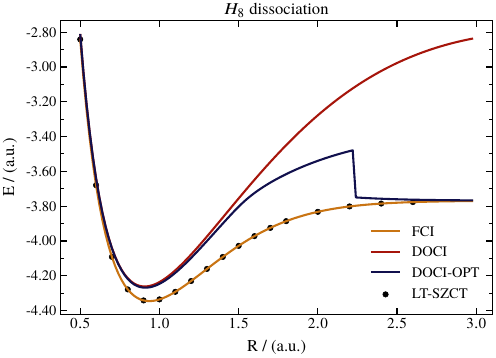} 
\caption{\label{fig1} Dissociation curve for the linear \ce{H8} chain in the STO-6G basis set as a function of nearest-neighbor distance. The full configuration interaction (FCI) reference is compared to the doubly-occupied configuration interaction (DOCI) seniority-zero wavefunction, with and without orbital optimization (OPT). The late-truncation seniority-zero canonical transformation approach presented in the paper is labeled LT-SZCT.}
\end{figure}

As a prototypical example of strong correlation, we start by considering the dissociation of a linear \ce{H8} chain in the STO-6G minimal basis, using orbital-optimized DOCI (DOCI-OPT) reference wave function. DOCI is known to be qualitatively correct along the dissociation curve and exact in the dissociation limit; see Figure \ref{fig1}. Orbital optimization is critical: DOCI without orbital optimization is quite poor except near equilibrium. 

\begin{figure}[h!]
\includegraphics[width=0.7\linewidth]{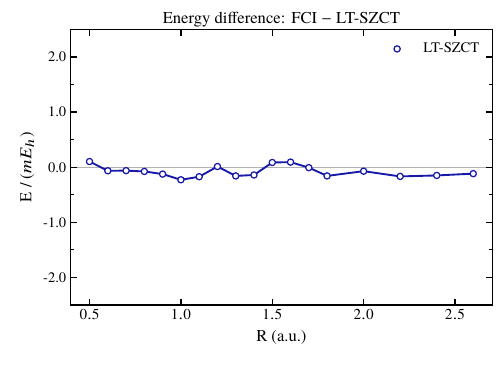} 
\caption{\label{fig2}  The energy difference between the LT-SZCT and FCI, in $mE_h$, for the dissociation of the linear \ce{H8} chain in the STO-6G basis set. The lack of variational character indicates that there are (small) errors associated with the approximation of the BCH expansion.}
\end{figure}

Our late-truncation seniority-zero canonical transformation (LT-SZCT) approach qualitatively (Figure \ref{fig1}) and quantitatively (Figure \ref{fig2}) reproduces the full configuration interaction reference. For compressed and near equilibrium distances, we observed (see Figure S1 in the supplementary material) that non-orbital-optimized DOCI also sufficed as a reference wavefunction, maintaining errors on the order of $10^{-4}mE_h$; however, orbital optimization is critical as one approaches the dissociation limit.

In general, the LT-SZCT is impressively accurate, with an average absolute error of $0.1078~mE_h$. From Figure \ref{fig2} we observe that the energy errors are not always negative, meaning that energy predictions from LT-SZCT are not upper bounds to the exact energy. This arises from the approximate evaluation of the higher-order terms in the BCH expansion in Eq. \ref{eq5} using operator decomposition. For the points where the LT-SZCT energy is smaller than the FCI result, the constraint on the norm  of the generator $A$ did not keep the transformation unitary, so the ground state energy of the transformed Hamiltonian $H_{SZ}$ was not bounded below by the FCI energy. Although the errors are very small, the variational character of the method can be restored by simply constraining the norm of the generator to be smaller. Finally, we identify two possible reasons for the remaining small jumps in the potential energy curves: First, during the optimization, the method might be shifting between (almost energetically degenerate) local minima. Second, given that the constraint over the generator was not determined with a general analytical expressions but intuitively based on the difference between the reference DOCI calculation and FCI, small deviations in the chosen constraint on the norm of the generator might induce jumps in the potential energy curve.

\subsection{\ce{BH}} 
\begin{figure}[h!]
\includegraphics[width=0.7\linewidth]{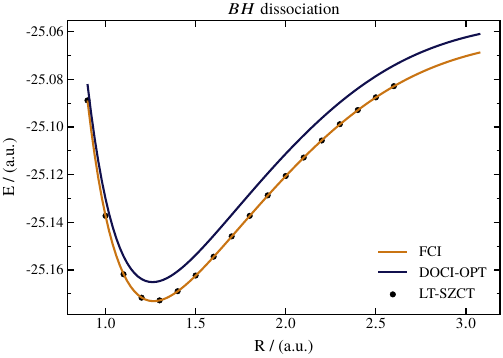} 
\caption{\label{fig3}\ce{BH} dissociation energy curve in the 6-31G basis set.}
\end{figure}

Next we consider the dissociation of \ce{BH}, examining two basis sets: 6-31G and cc-pVDZ. As has been previously reported,\cite{Limacher19032014} DOCI-OPT provides a qualitatively and quantitatively reliable description of the \ce{BH} dissociation in modest basis sets; for 6-31G, the mean deviation from FCI is $\approx 9~mE_h$ (cf. Figure \ref{fig3}). This level of accuracy makes \ce{BH} a convenient test system for assessing post-seniority-zero corrections under conditions where the seniority-zero reference is already close to the exact result.

\begin{figure}[h!]
\includegraphics[width=0.7\linewidth]{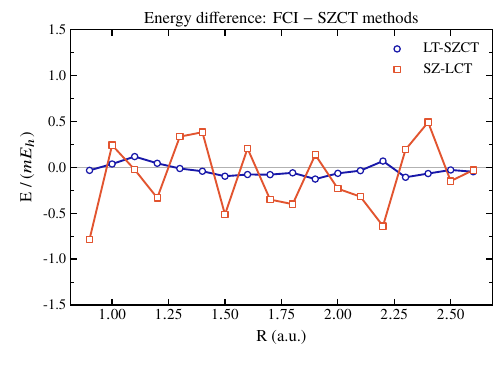} 
\caption{\label{fig4} Energy deviations relative to FCI for SZ-LCT and LT-SZCT, in $mE_h$, for dissociating \ce{BH} dissociation in the 6-31G basis set.}
\end{figure}

\begin{figure}[h!]
\includegraphics[width=0.7\linewidth]{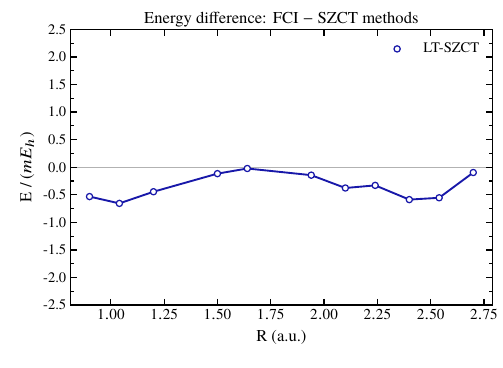} 
\caption{\label{fig4.1} Energy deviations relative to FCI  LT-SZCT, in $mE_h$, for dissociating \ce{BH} dissociation in the cc-pVDZ basis set.}
\end{figure}

In this regime, the generator $A$ can be kept relatively small. To isolate the effect of late truncation (i.e., treating the first two terms in Eq. (\ref{eq4}) exactly), we compare LT-SZCT to seniority-zero linear canonical transformation theory (SZ-LCT), in which all commutators in the BCH expansion are truncated to two-body operators.\cite{caleroosorio2025seniorityzerolinearcanonicaltransformation}. Even for this ``easy'' case, LT-SZCT is a remarkable improvement over SZ-LCT, with average error of $6.46 \times 10^{-5}~ E_h$, surpassing SZ-LCT in accuracy and stability. By delaying the truncation of the BCH expansion, LT-SZCT seems to mitigate the local-minimum problems that give SZ-LCT jagged behavior in the energy deviations (cf. Figure \ref{fig4}). As both methods have the same effective computational scaling, LT-SZCT is to be preferred.

We next turn to the cc-pVDZ results, which provide a more stringent assessment of the methods in an enlarged orbital space. Although DOCI-OPT remains a qualitatively good representation of the FCI dissociation curve, its quantitative performance deteriorates, with a mean deviation of $\approx 22~mE_h$ (cf. Figure S2 in supplementary material). Consequently, larger generators $\hat{A}$ are required to recover the missing dynamic correlation. In Figure \ref{fig4.1} we report the energy deviations from FCI for our method. While the errors are larger than in the 6-31G basis, they remain on the order of $10^{-4}E_h$, providing evidence that the method remains accurate beyond the small-basis regime, where dynamic correlation effects are more fully expressed. In addition, LT-SZCT maintains its advantages over SZ-LCT, without the pronounced peaks in the energy differences that arise from local-minimum issues.

\subsection{\ce{N2}} 
\begin{figure}[h!]
\includegraphics[width=0.7\linewidth]{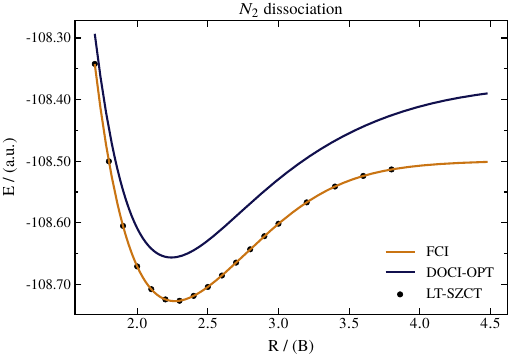} 
\caption{\label{fig5} \ce{N2} dissociation energy curve in STO-6G basis set.}
\end{figure}

Finally, we consider \ce{N2} dissociation in the STO-6G basis set. This is a more challenging test for the method because the performance of DOCI deteriorates when multiple bonds are broken.\cite{bytautasSeniorityNumberDescription2015} When breaking the triple bond in \ce{N2}, DOCI-OPT energy predictions have a stable error of about $8 \times10^{-2} ~ E_h$ for compressed and equilibrium geometries, reaching $\sim 10^{-1}~E_h$ in the dissociation limit (cf. figure \ref{fig5}).

\begin{figure}[h!]
\includegraphics[width=0.7\linewidth]{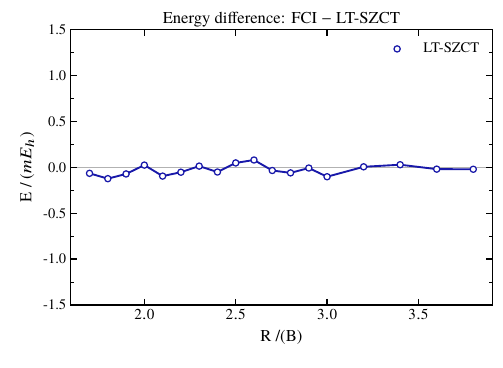} 
\caption{\label{fig6}  Energy difference between the LT-SZCT method and FCI in $mE_h$ for \ce{N2} dissociation in STO-6G basis set.}
\end{figure}
Remarkably, even though DOCI-OPT is \textit{quantitatively} inaccurate, LT-SZCT gives excellent performace, with small stable errors (the average absolute error is $5.07 \times 10^{-5}~ E_h$) along the whole dissociation curve (cf. figure \ref{fig6}). For stretched geometries ($R > 3.5 $ a.u.), the DOCI-OPT energy error is about $0.1 E_h$ and the LT-SZCT method improves the energy prediction by  4 orders of magnitude. 


\section{\label{conclusions}Conclusions}
We show how to add residual (dynamic) correlation to seniority-zero wave functions by constructing a unitary transformation of the Hamiltonian; the idea is to build a transformed Hamiltonian whose ground state is, as nearly as possible, the orbital-optimized doubly-occupied configuration interaction wave function (DOCI-OPT). Following the strategy of canonical transformation (CT) theory, we evaluate the unitary transformation using the Baker-Campbell-Hausdorff (BCH) expansion, exploiting the seniority-zero structure of the wavefunction to compute the first three terms (involving up to the 4-electron reduced density matrix (RDM)) exactly at a reasonable cost, and then using operator decomposition to approximate the third- and higher-order contributions in the BCH expansion with 2-body operators. With this scheme, the truncation error is of order $\mathcal{O}\left(\lVert \hat{A}\rVert^3\right)$ and, therefore, if the norm of the generator of the unitary transformation is kept small, the approximate BCH expansion accurately approximates the targeted unitary transformation. Our computational implementation leverages the advantages of the seniority-zero reference and parallelization.

In general, the method delivered highly accurate results for all tested molecules, with mean absolute errors in the order of $10^{-4}E_h$. We thought that results might be further improved by including 3-body and/or 4-body terms in the operator decomposition. E.g., instead of approximating the third- and higher-order terms in the BCH with 2-body operators (cf. Eq. (\ref{eq5})), we can approximate them with 3-body operators,
\begin{equation}
\begin{aligned}
e^{-\hat{A}}\hat{H}e^{\hat{A}} = \hat{H} &+ \left[\hat{H},\hat{A}\right] + \frac{1}{2!} \left[\left[\hat{H},\hat{A}\right],\hat{A}\right]  \\[8pt] &+ \frac{1}{3!} \left[\left[\left[\hat{H},\hat{A}\right],\hat{A}\right]_{1,2,3}, \hat{A}\right]_{1,2,3} +...~.
\end{aligned}
    \label{eq18}
\end{equation}
This is theoretically more accurate than the 2-body truncation, however it decreases the computational efficiency significantly because: (1) evaluating the truncated commutator $\left[\left[\hat{H},\hat{A}\right],\hat{A}\right]_{1,2,3}$ scales as $\mathcal{O}(N^7)$, while truncating at the 2-body level scales only as $\mathcal{O}(N^5)$. (2) The recursive evaluation of the commutators with this truncation requires storing and manipulating the three-body operator. The additional cost of including 4-body terms would be higher still. We implemented Eq. (\ref{eq18}) and tested it for \ce{H8}. The results were very similar to the 2-body truncation (Eq. (\ref{eq5})), showing no advantage in accuracy but a significant increase in computational cost. We believe, therefore, that our commutator truncation choice provides the best balance between accuracy and performance.

Comparison between LT-SZCT and the previously introduced SZ-LCT show that the late-truncation strategy decreases errors by an order of magnitude and resolves erratic oscillations around the true solution (presumably due to the proliferation of local minima in the SZ-LCT objective function), while the computational scaling is identical. LT-SZCT even works when DOCI-OPT gives errors up to $.1~E_h$, as in multiple-bond dissociation. 

In light of recent results indicating that seniority-zero wavefunctions can always be expressed as (number-symmetry-broken-and-restored) antisymmetric products of interacting geminals, one can interpret this method as a traditional approach for adding dynamic correlation to a (quasiparticle) mean-field.\cite{martinezgonzalezSeniorityzeroStatesAre2025} It is interesting to contemplate whether the late-truncation canonical-transformation approach might be feasible for other sorts of quasiparticle-mean-field wavefunctions also.\cite{zobokiCompositeParticlesQuantum2013b,martinezgonzalez2025TCA} Our current work is focused on three main aspects: (1) extending the method/algorithm to automatically determine the maximum norm of the generator $A$ that keeps the transformation (approximately) unitary. This can be done by comparing the magnitude of the last exactly evaluated term $\left[\left[H,A\right],A\right]$ with the first approximated term $\left[\left[\left[H,A\right]_{1,2},A\right]_{1,2},A\right]_{1,2}$, (2) developing a more efficient algorithm for evaluating the gradient of the objective function, and (3) applying this strategy to low-cost seniority-zero wavefunction ans\"{a}tze (e.g., from geminal-based approaches), thereby avoiding the costly DOCI-OPT calculation.\cite{surjanIntroductionTheoryGeminals1999,10.1063/5.0088602,limacher2016new,richerGraphicalApproachInterpreting2025,kimFlexibleAnsatzNbody2021} 

\begin{acknowledgement}
The authors acknowledge support from the Canada Research Chairs (CRC-2022-00196), NSERC (Discovery RGPIN/06707-2024 and Alliance ALLRP/592521-2023), and the Digital Research Alliance of Canada. 
\end{acknowledgement}

\begin{suppinfo}
The data that support the findings of this study are available from the corresponding author upon reasonable request. Tabulated data supporting the figures is included as supplementary material.
\end{suppinfo}

\bibliography{achemso}

\end{document}